\documentclass[useAMS,usenatbib]{mn2e}
\usepackage{amssymb,amsmath,epsfig,times, natbib}
\title[Centaurus gas motions]{Constraining gas motions in the Centaurus cluster 
using X-ray surface brightness fluctuations and metal diffusion}
\author[S. A. Walker et al.]{S. A. Walker,$^1$\thanks{Email: 
    swalker@ast.cam.ac.uk} J. S. Sanders$^2$ and A. C. Fabian$^1$ \\
  $^1$Institute of Astronomy, Madingley Road, Cambridge CB3 0HA \\
  $^2$Max-Planck-Institute fur extraterrestrische Physik, 85748 Garching, 
Germany  \\
  \\
    \\
   \\
   \\
}
\date{}

\begin{document}

\maketitle

\begin{abstract}
We compare two different methods of constraining the characteristic velocity and 
spatial scales of gas motions in the X-ray bright, nearby Centaurus cluster, 
using new deep (760ks) Chandra observations. The power spectrum of excess 
surface brightness fluctuations in the 0.5-6.0 keV band in a sector to the west is 
measured and compared to theoretical expectations for Kolmogorov index 
fluctuations. The observed power spectrum is flatter than these expectations, 
and the surface brightness fluctuations are around the 8 percent level on length 
scales of 2 kpc. We convert the 2D power spectrum of fluctuations into a 3D 
power spectrum using the method of Churazov et al., and then convert this into 
constraints on the one-component velocity of the gas motions as a function of 
their length scale. We find one-component velocities in the range 100-150 km/s 
on spatial scales of 4-10 kpc. An independent constraint on the characteristic 
velocity and length scales of the gas motions is then found by considering the 
diffusion coefficient needed to explain the distribution of metals in the 
Centaurus cluster, combined with the need to balance the rate of gas cooling 
with the rate of heat dissipated by the gas motions. We find that these two 
methods of constraining the velocity and length scales of the gas motions are in 
good agreement.   
 
\end{abstract}


\section{Introduction}
In the centres of galaxy clusters the cooling time of gas is much shorter than 
the Hubble time. Feedback from the central AGN is needed to counteract this 
cooling to prevent the formation of a cooling flow and limit the rate of star 
formation. The precise way in which the AGN feedback energy is dissipated into 
the intracluster medium (ICM), and how this can so accurately balance the rate 
of cooling, remains under debate. 

Turbulence is one mechanism through which the energy from the central AGN could 
be dissipated into the ICM to counteract cooling. Direct constraints on the 
level of turbulence in the cores of galaxy clusters remain challenging with 
existing X-ray instruments. Measuring the width of emission lines seen with the 
RGS (reflection grating spectrometer) instrument on \emph{XMM-Newton}  can 
provide upper limits on the level of gas motions. \citet{Sanders2010} used this 
method to find a limit of 274 km s$^{-1}$ for random gas motions in the core of 
Abell 1835, while in the sample of clusters studied in \citet{Sanders2011} 15 
clusters were found for which the upper limits gave a turbulence to thermal 
energy density ratio of less than 20 percent.

Recent high spatial resolution studies of nearby clusters have investigated 
spatial fluctuations in the X-ray emission from the centres of galaxy clusters, 
which can place limits on the level of gas motions in the ICM due to turbulence. 
\citet{ChurazovComa} studied the X-ray surface brightness fluctuations in the 
core of the non-cool core Coma cluster with $Chandra$ and \emph{XMM-Newton} 
data, using the $\Delta$-variance technique (\citealt{Deltavariance}) to 
characterise the magnitude of fluctuations as a function of their spatial scale. 
They found that the characteristic amplitude of excess density fluctuations 
varied from $\sim$5 percent on scales of $\sim$30 kpc, up to 7-10 per cent on 
much larger scales of $\sim$ 500 kpc.

\citet{SandersAWM7} investigated the X-ray surface brightness fluctuations in 
the cool-core galaxy cluster AWM7, and found implied 3D density density 
fluctuations with a standard deviation of around 4 percent, with 3D pressure 
variations of at most 4 per cent as well. They compared the shape of the power 
spectrum of excess surface brightness fluctuations with that expected from 
Kolmogorov index fluctuations, and found the observed spectrum to be flatter 
than these expectations. By comparing the northern and southern sectors of AWM7, 
they found significant spatial variations in the fluctuations. In the northern 
sector studied, there was a factor of 2 less $\Delta$-variance relative to the 
southern sector studied.
  
Recently, \citet{Zhuravleva2014} and \citet{Zhuravleva2015} have studied the 
X-ray surface brightness fluctuations in the Perseus and Virgo clusters. In 
\citet{Zhuravleva2014} the 2D power spectrum of surface brightness fluctuations 
is converted into a 3D power spectrum, which is subsequently converted into the 
one-component velocities of the gas motions needed to cause the fluctuations, 
which are found as a function of their spatial size. This method was applied to 
annuli of increasing radius, and the characteristic one-component velocity was 
found to decrease from around 200 km/s in the central 30 kpc to 100 km/s at a 
distance of 200 kpc from the centre. They then calculated the turbulent heating 
rate expected for gas motions with the velocity and spatial sizes found. For the 
Perseus and Virgo clusters, they find that the rate of turbulent heating deduced 
using this method is roughly equal to the cooling rate of the gas in all of the 
radial annuli studied, adding support to the idea that turbulence plays a role 
in balancing gas cooling in cluster cores.

Independent constraints on the velocity and spatial scales of the gas motions 
can also be found from the distribution of metals in the abundance peak around 
the brightest cluster galaxy, BCG  (\citealt{Rebusco2005}, 
\citealt{Rebusco2006}). This method assumes that, as the metals are continuously deposited into the ICM from stars in the BCG over the life time of the cluster, the profile of the metal abundance 
from this deposition follows the light profile of the BCG. The metals then diffuse outwards to 
form the much broader metal abundance peaks in cluster cores that we see today. 
By assuming that metal diffusion has proceeded through stochastic gas motions, 
this allows the metal diffusion coefficient, $D$ to be constrained, which is 
proportional to the product of the characteristic velocity, $v$, and length 
scales, $l$,  of these gas motions. If it is then assumed that the heating rate 
from the dissipation of these stochastic gas motions balances the gas cooling 
rate, $v$ and $l$ can be constrained.  \citealt{Rebusco2005} used this technique 
for the Perseus cluster and found that a diffusion coefficient of 
$2\times10^{29}$ cm$^{2}$ s$^{-1}$ is needed to explain the observed abundance 
profile. They deduced that the turbulent gas motions in Perseus have velocities 
of $\sim$300 km s$^{-1}$ on spatial scales of 20 kpc.

Here we investigate new, deep (760 ks) Chandra observations of the cool-core 
Centaurus cluster (P.I J.S. Sanders) to compare the constraints on the 
characteristic velocity and spatial scales of gas motions obtained by 
considering the X-ray surface brightness fluctuations and the distribution of 
metals in the ICM. A detailed study of these new Chandra observations 
will be presented in Sanders et al. (submitted). Centaurus' proximity (it 
is the second nearest galaxy cluster), combined with its high X-ray flux makes 
it an ideal target for exploring X-ray surface brightness fluctuations. 

\citet{Graham2006} have previously used the existing 200ks Chandra observation 
to investigate metal diffusion following the method of \citet{Rebusco2005}. The 
azimuthally averaged metal abundance profile for Centaurus has a prominent drop 
between 20-50 kpc from the core which requires a comparatively low diffusion 
coefficient to maintain. \citet{Graham2006} found a best fitting diffusion 
coefficient in the range 2-4$\times10^{28}$ cm$^{2}$ s$^{-1}$, nearly an order 
of magnitude lower than that found for Perseus in \citet{Rebusco2005}. The very 
high metallicity around the BCG of Centaurus, which peaks at nearly 3 
Z$_{\odot}$ (\citealt{Fabian2005}), makes it an excellent target for applying 
this method.

We use a standard $\Lambda$CDM cosmology with $H_{0}=70$  km s$^{-1}$
Mpc$^{-1}$, $\Omega_{M}=0.3$, $\Omega_{\Lambda}$=0.7. For the redshift of the 
Centaurus cluster, $z=0.011$, the luminosity distance is 49.5 Mpc and the 
angular scale is 0.235 kpc/arcsec. All errors unless
otherwise stated are at the 1 $\sigma$ level. 

\begin{figure}
  \begin{center}
    \leavevmode
    
      \epsfig{figure=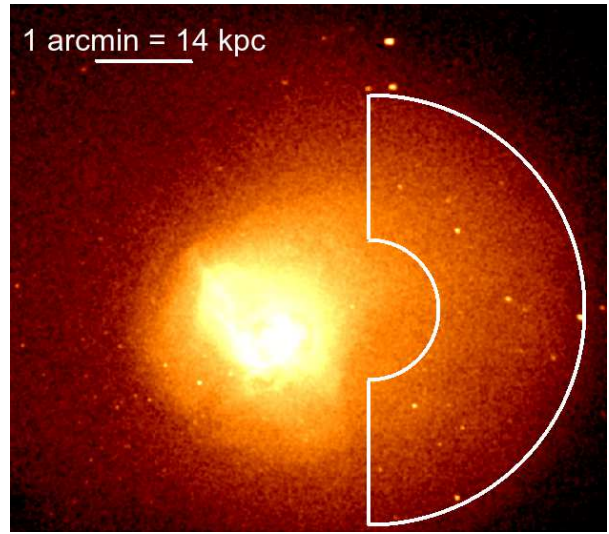,
        width=0.95\linewidth}

      \caption{Exposure corrected 0.5-6.0 keV mosaic image of the Centaurus 
cluster. The white sector region shown is used for the analysis of the surface 
brightness fluctuations. Point sources have not been removed from this image. }
      \label{Cen_image}
  \end{center}
\end{figure}

\section{Data}

The new deep Chandra observations of Centaurus will be presented and explored in 
depth in Sanders et al. (submitted). The observations used are tabulated in 
table \ref{obsids}. The data were reprocessed using CIAO v4.6. Lightcurves for each observation were examined, with any periods deviating from the standard deviation by more than 2.5 $\sigma$ removed, leaving a total clean exposure of 760ks.

\begin{table}
\begin{center}
\caption{Chandra observations used}
\label{obsids}
\leavevmode
\begin{tabular}{ lll } \hline \hline
Obsid & Cleaned exposure/ ks & Date \\ \hline
504 & 26 & 2000-05-22\\
505 & 10 & 2000-06-08\\
4190 & 34.1 & 2003-04-18\\
4191 & 33.8 & 2003-04-18\\
4954 & 85.4 & 2004-04-01\\
4955 &43.1 & 2004-04-02\\
5310 & 48.7 & 2004-04-04\\
16223 & 176.2 & 2014-05-26\\
16224 &41.2 & 2014-04-09\\
16225 & 29.7 & 2014-04-26\\
16534 & 55 & 2014-06-05\\
16607 & 44.6 & 2014-04-12 \\
16608 &33.3 & 2014-04-07\\
16609 &81.7 & 2014-05-04\\
16610 &17.1 & 2014-04-27\\ \hline
Total & 760 & \\ \hline
\end{tabular}
\end{center}
\end{table}

\section{Analysis}

\begin{figure*}
  \begin{center}
    \leavevmode
    \hbox{
      \epsfig{figure=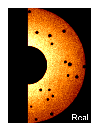,
        width=0.45\linewidth}
        \epsfig{figure=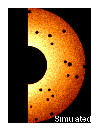,
        width=0.45\linewidth}      
            }
    
      \caption{Left: Actual 0.5-6.0keV band image from the sector region shown 
in Fig. \ref{Cen_image}, with point sources removed. Right: Simulated 0.5-6.0keV 
band image from the same sector, obtained by modelling the surface brightness of 
this region with 10 ellipsoids. }
      \label{real_v_sim_sector}
  \end{center}
\end{figure*}

The exposure corrected and background subtracted 0.5-6.0 keV band stacked image 
of Centaurus is shown in Fig. \ref{Cen_image}. The stacked, exposure corrected image was produced following the methods described in \citet{Walker2014_A1795}. To study the surface brightness 
fluctuations we investigate the region shown by the white sector, which spans a 
radial range of 23-43 kpc from the centre of the BCG. This region lies between 
the core and the large western cold front. It has a high X-ray surface 
brightness over a large area and is relatively smooth in its X-ray emission. It 
was chosen to avoid any surface brightness features which are obviously due to 
feedback from the core (i.e. the cavities and filaments) or due to gas sloshing 
(i.e. the eastern and western cold fronts and the `bay' feature to the south 
west of the core).

The left hand panel of Fig. \ref{real_v_sim_sector} shows the sector region by 
itself, with point sources removed. Point sources were identified by the tool 
wavdetect, using a range of wavelet sizes between 1 and 16 pixels. The large 
scale surface brightness distribution in this region was modelled by fitting ten 
ellipsoids to it, producing a smooth model image. This ten ellipsoid model was used because it was found to be the simplest way of accurately modelling the large scale surface brightness distribution of the region. The power spectrum of this model on large scales matches that of the real observations very closely. This smooth model image was 
multiplied by the exposure map, and then a simulated image in the 0.5-6.0 keV 
band was obtained by making a Poisson realisation. This was then exposure 
corrected by dividing by the exposure map, producing the simulated image shown 
in the right hand panel of Fig. \ref{real_v_sim_sector}.

\begin{figure*}
  \begin{center}
    \leavevmode
    \hbox{
      \epsfig{figure=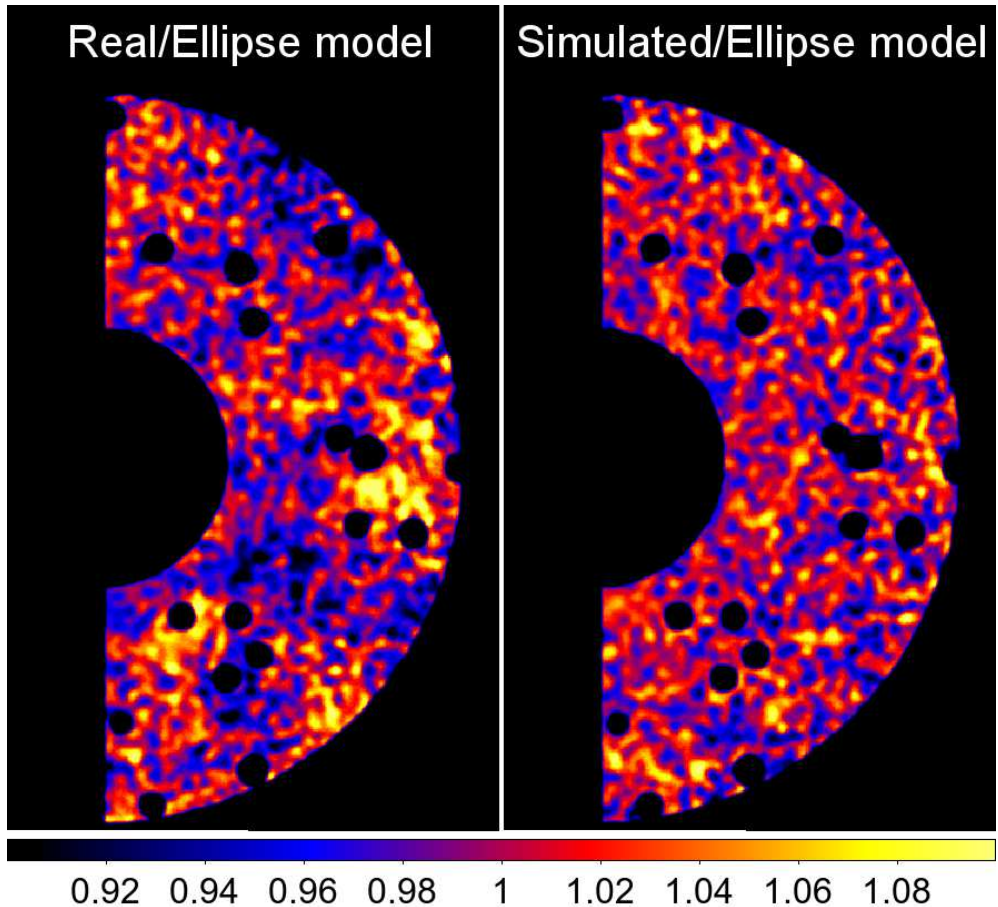,
        width=0.75\linewidth}
     
            }
    
      \caption{Left: Ratio of the actual data to the best fitting 10 ellipsoidal 
model. Right: Ratio of the simulated data to the best fitting 10 ellipsoidal 
model. }
      \label{real_v_sim_ratio_sector}
  \end{center}
\end{figure*}

In Fig. \ref{real_v_sim_ratio_sector} we show the ratio of the real image (left) 
and the simulated image (right) to the smooth image we obtained by fitting 10 
ellipsoids to this region. There are clear structures in the residuals of the 
real data to this smooth model (left hand panel) at the 10 percent level.

\section{Comparing fluctuation power spectra}

To compare the shape and magnitude of the observed power spectrum of surface 
brightness fluctuations with theoretical expectations for hydrodynamical 
turbulence, we follow the approach of \citet{SandersAWM7}. We use the simulated 
image of the sector, obtained by fitting 10 ellipsoids to the region, and add to 
it surface brightness fluctuations with a powerlaw power-spectrum. These surface 
brightness fluctuations are generated by finding the fast Fourier transform of a 
cube of dimensions 700$\times$700$\times$700, which is larger than the size of 
the region of the Centaurus image explored (in terms of pixels). Each pixel in 
the cube was scaled by a powerlaw power spectrum of index -11/3 (the shape 
expected for the electron density power spectrum for a Kolmogorov process), 
using a normal distribution for the real and imaginary components. Following the 
inverse transform, the real component for each pixel was taken. The 
powerspectrum was truncated to remove structure on scales larger than 30 kpc 
(the radial range spanned by the sector examined). We ensured that this truncation scale is consistent with the use of the 10 ellipsoids model for removing large-scale perturbations.

We deproject the surface brightness profile from the Centaurus observations to 
obtain the deprojected emissivity model, $E(x,y,z)$. We then integrate the cubic 
fluctuation, $F(x,y,z)$ (normalised to have a standard deviation of 1), field 
along the $z$-axis, multiplying by the deprojected emissivity profile and a 
scale factor determining the magnitude of input fluctuations, $\epsilon$,

\begin{eqnarray}
S(x,y) = \int [1 + \epsilon F(x,y,z)]E(x,y,z) {\rm d}z
\label{project_fluctations}
\end{eqnarray}

from which we obtain a simulated image, $S(x,y)$ containing powerlaw 
fluctuations of magnitude 4, 8 and 16 percent in surface brightness. The Chandra 
exposure maps were then used to create Poisson realisations of these images, 
allowing us to compare these to the actual observations.

\begin{figure*}
  \begin{center}
    \leavevmode
    \hbox{
      \epsfig{figure=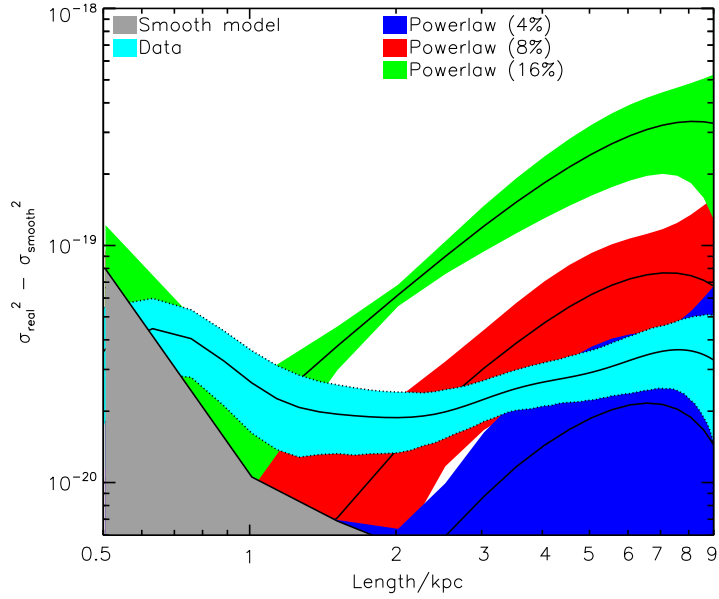,
        width=0.6\linewidth}
     
            }
    
      \caption{The power spectrum for the excess surface brightness fluctuations 
in the real data, from which the spectrum of the smooth model has been 
subtracted, is shown as the cyan region. This is compared to the power spectra 
of simulated images to which 4, 8 and 16 percent Kolmogorov index fluctuations 
have been added.     }
      \label{powerspectra}
  \end{center}
\end{figure*}

We then used the delta variance technique to find the power spectra of the 
simulated images to which fluctuations of magnitude 4, 8 and 16 percent in 
surface brightness have been added. These magnitudes correspond to the RMS of the surface brightness over the image. These power spectra are shown in Fig. 
\ref{powerspectra}. The power spectrum of the simulated ellipsoidal model with 
zero fluctuations has been subtracted from these power spectra, allowing us to 
see the excess power that is just due to the fluctuations we have added. The 
error ranges of the power spectra were calculated by simulating 1000 Poisson 
realisations of each image and finding the standard deviation of the power 
spectra from these. We performed tests to ensure the results are not sensitive to variations in the procedure used. The analysis was repeated but this time the model was subtracted from the data first, and then the power spectrum was found. The results were found to be the same as those obtained by subtracting the power spectrum of the model from that of the data. 

The power spectrum of the actual data (shown as the cyan region in Fig. 
\ref{powerspectra}) is then compared to these simulations of different 
fluctuation levels. We see that the power spectrum of the real data is flatter 
in shape than the spectra made assuming fluctuations using a standard Kolmogorov 
powerlaw index of -11/3. This is similar to what was found by 
\citet{SandersAWM7} for AWM7. It may be possible that unresolved compact sources contribute to the apparent flatness of the observed spectrum. The 10 ellipsoidal model used only has a noticeable affect on scales larger than around 10 kpc, so it is unlikely that the use of this model contributes to the observed flattened shape.

\section{Velocity power spectra}

\begin{figure*}
  \begin{center}
    \leavevmode
    \hbox{
      \epsfig{figure=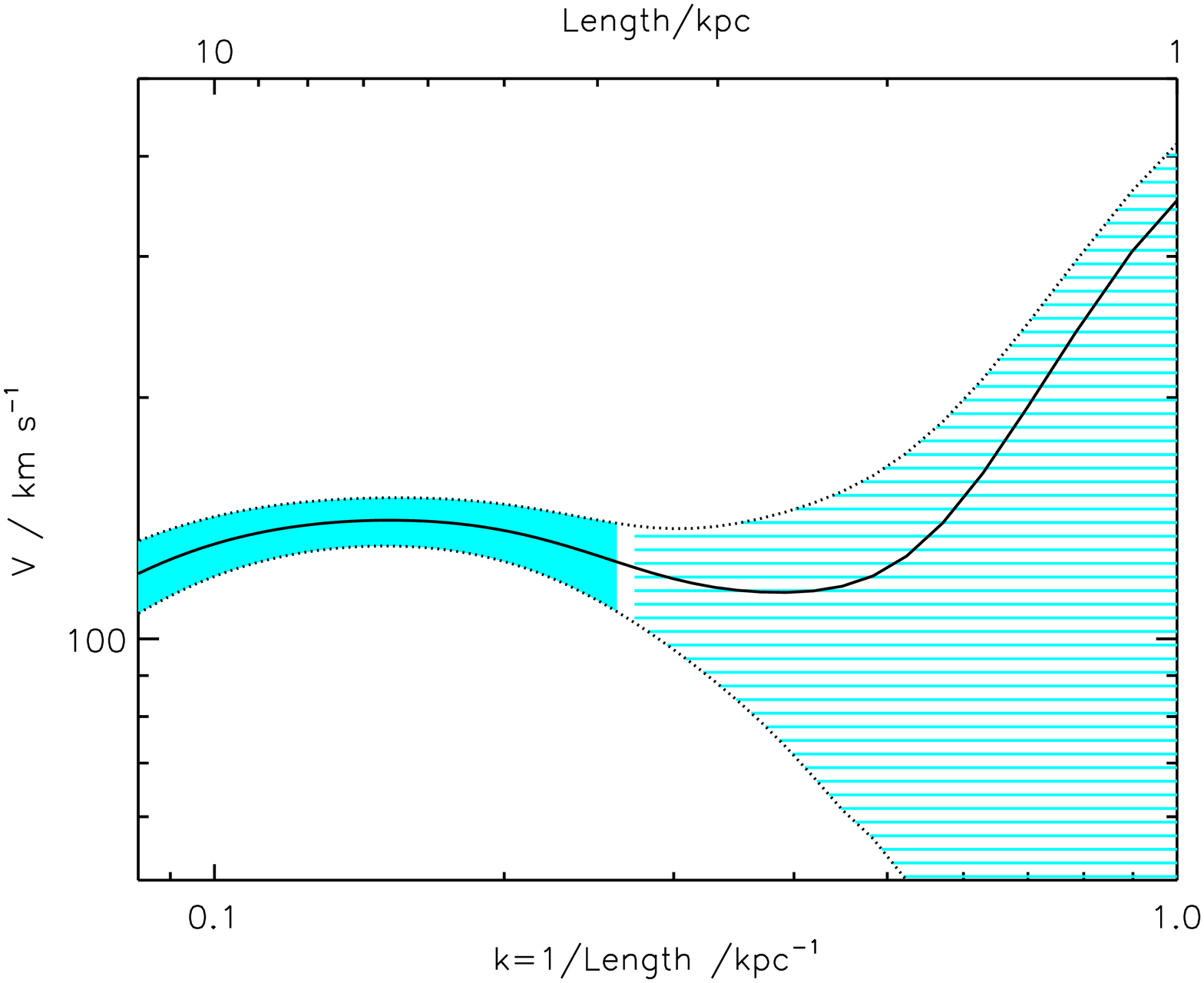,
        width=0.5\linewidth}
          \epsfig{figure=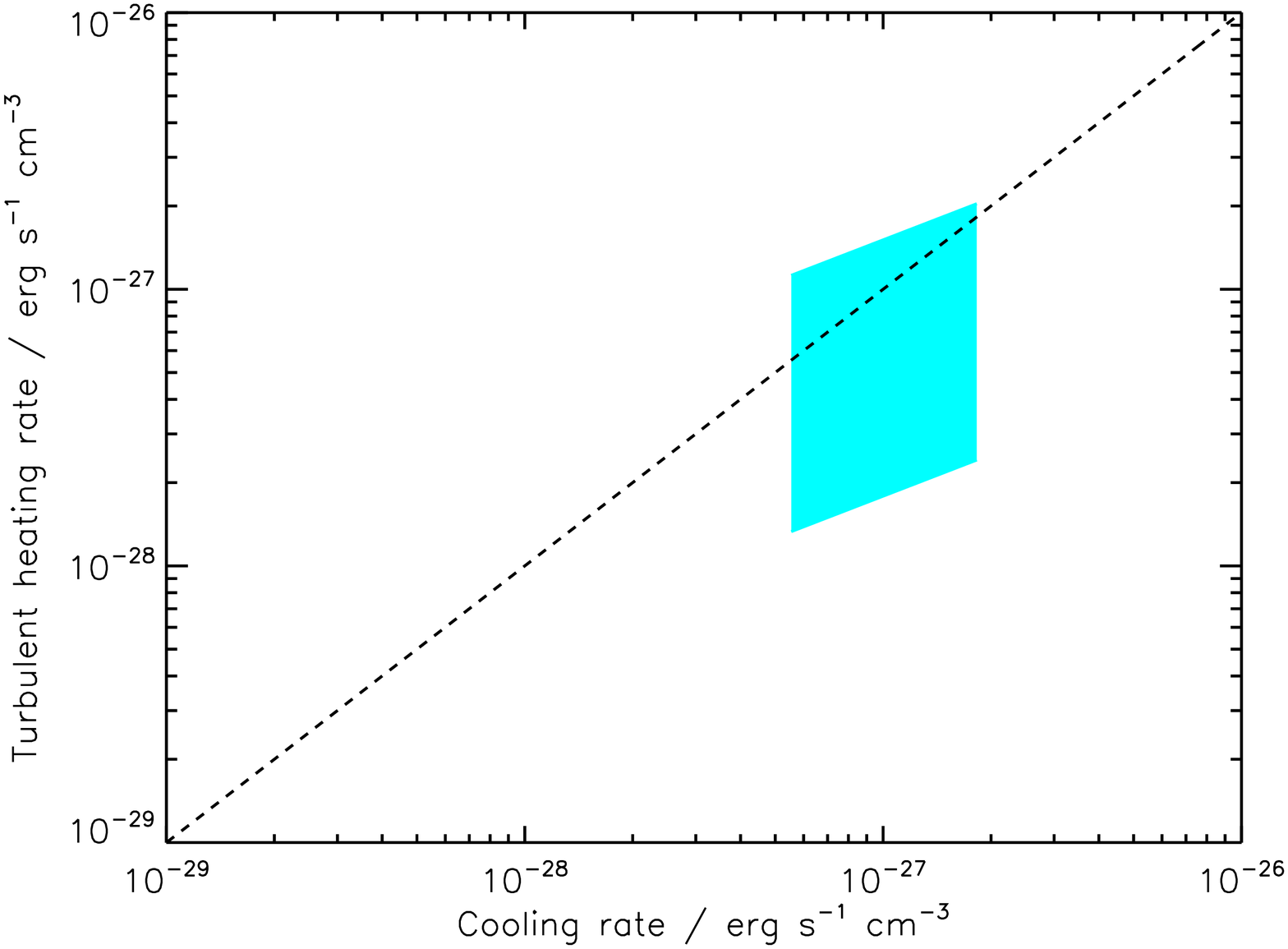,
        width=0.5\linewidth}
            }
    
      \caption{\emph{Left:} Constraints on the one component velocity of the gas 
motions as a function of their wavenumber, derived from the power spectrum of 
excess surface brightness fluctuations. The solid cyan region shows the region 
where systematic and statistic uncertainties least affect the measurement. 
\emph{Right:} The cyan region shows the turbulent heating rate inferred from the 
surface brightness fluctuations against the cooling rate of the sector region 
studied. This agrees reasonably well with equilibrium between heating and 
cooling, which is shown by the dashed black line.   }
      \label{velocitypowerspectra}
  \end{center}
\end{figure*}

Following the method detailed in \citet{ChurazovComa}, we can convert the 2D 
power spectrum obtained from our image analysis, $P_{\rm 2D}$, into a 3D power 
spectrum $P_{\rm 3D}$ using

\begin{eqnarray}
P_{2D}(k)\approx 4P_{3D}(k)\int{|W(k_z)|^2dk_z},
\label{eq:klarge}
\end{eqnarray}
where $|W(k_z)|^2$ is the 1D power spectrum of the normalised emissivity, 
$\eta(z)$, along the line of sight

\begin{eqnarray}
W(k_z)=\int \eta(z) e^{-i2\pi z k_z} dz.
\label{eq:w}
\end{eqnarray}
where $\eta(z)$ is related to the density distribution $n^2_{0}(x,y,z)$ by

\begin{eqnarray}
\eta(x,y,z)=\frac{n^2_{0}(x,y,z)}{\int
  n^2_{0}(x,y,z')dz'}=\frac{n^2_{0}(x,y,z)}{I_0(x,y)}.
\end{eqnarray} 
and for the small area of the cluster we consider the dependence on $x$ and $y$ 
can be neglected, such that 
$\eta(x,y,z) \approx \eta(z)$. 

We can then follow the approach of \citet{Zhuravleva2014} and 
\citet{Zhuravleva2015} to convert this 3D power spectrum into the one-component 
velocity, $V_{1,k}$ power spectrum. For the stratified atmospheres of relaxed 
galaxy clusters, and within the inertial range of scales, the one-component 
velocity is expected to be proportional to the amplitude of density 
fluctuations, $A_{3D}= \frac{\delta \rho_{k}}{\rho_{0}}$ at each length scale 
$l=1/k$, so that we can write

\begin{eqnarray}
A_{3D} = \frac{\delta \rho_{k}}{\rho_{0}} = \eta_{1} \frac{V_{1,k}}{c_s}
\end{eqnarray} 
where $c_s$ is the sound speed. \citet{Zhuravleva2014L} found using cosmological 
simulations of galaxy clusters that the average value of the proportionality 
constant $\eta_{1}$ is consistent with unity ($\eta_{1} = 1 \pm 0.3$). 

The amplitude of the density fluctuations is related to the 3D power spectrum by

\begin{eqnarray}
A_{3D} = \sqrt{P_{3D}(k) 4 \pi k^{3}}
\end{eqnarray} 
which allows us to find the amplitude of the one-component velocity of the gas 
motions as a function of wavenumber ($k=1/l$) from

\begin{eqnarray}
V_{1,k} \approx c_s \sqrt{P_{3D}(k) 4 \pi k^{3}} / \eta_{1}
\end{eqnarray} 

The resulting plot of $V_{1,k}$ against wavenumber is shown in the left hand 
panel of Fig. \ref{velocitypowerspectra}, where the solid cyan region shows 
where the constraint is least affected by systematic and statistic 
uncertainties. 

In the right hand panel of Fig. \ref{velocitypowerspectra} we show the turbulent 
heating rate for the sector derived from the surface brightness fluctuations 
using equation \ref{heating_rate_turbulence} plotted against the cooling rate 
for the same region. The two appear to agree reasonably well. This is similar to 
the findings for the Perseus and Virgo clusters studied in 
\citet{Zhuravleva2014}.

\section{Constraints on velocity and spatial scales considering metal diffusion 
thermal equilibrium}

\begin{figure*}
  \begin{center}
    \leavevmode
    \hbox{
      \epsfig{figure=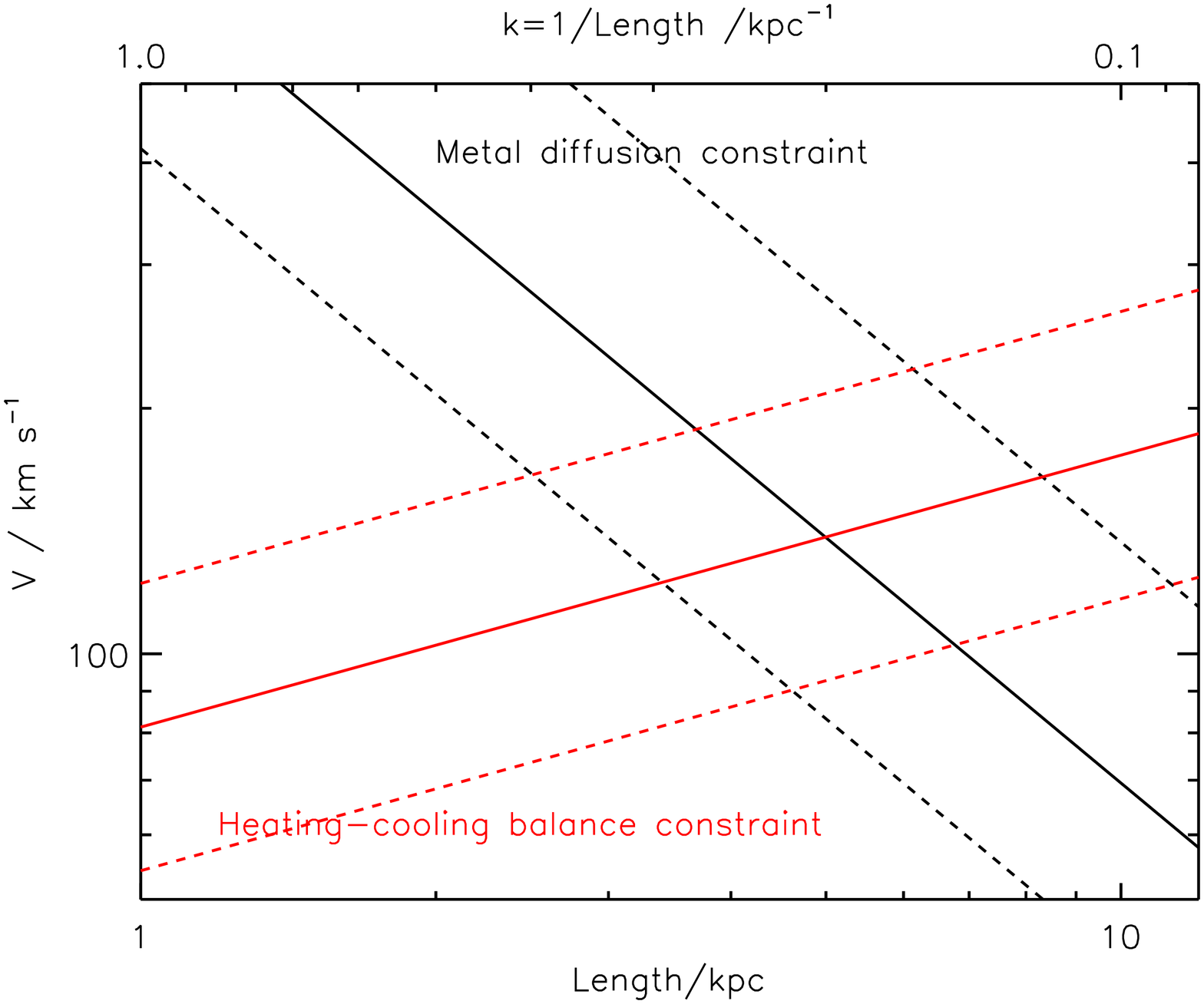,
        width=0.5\linewidth}
         \epsfig{figure=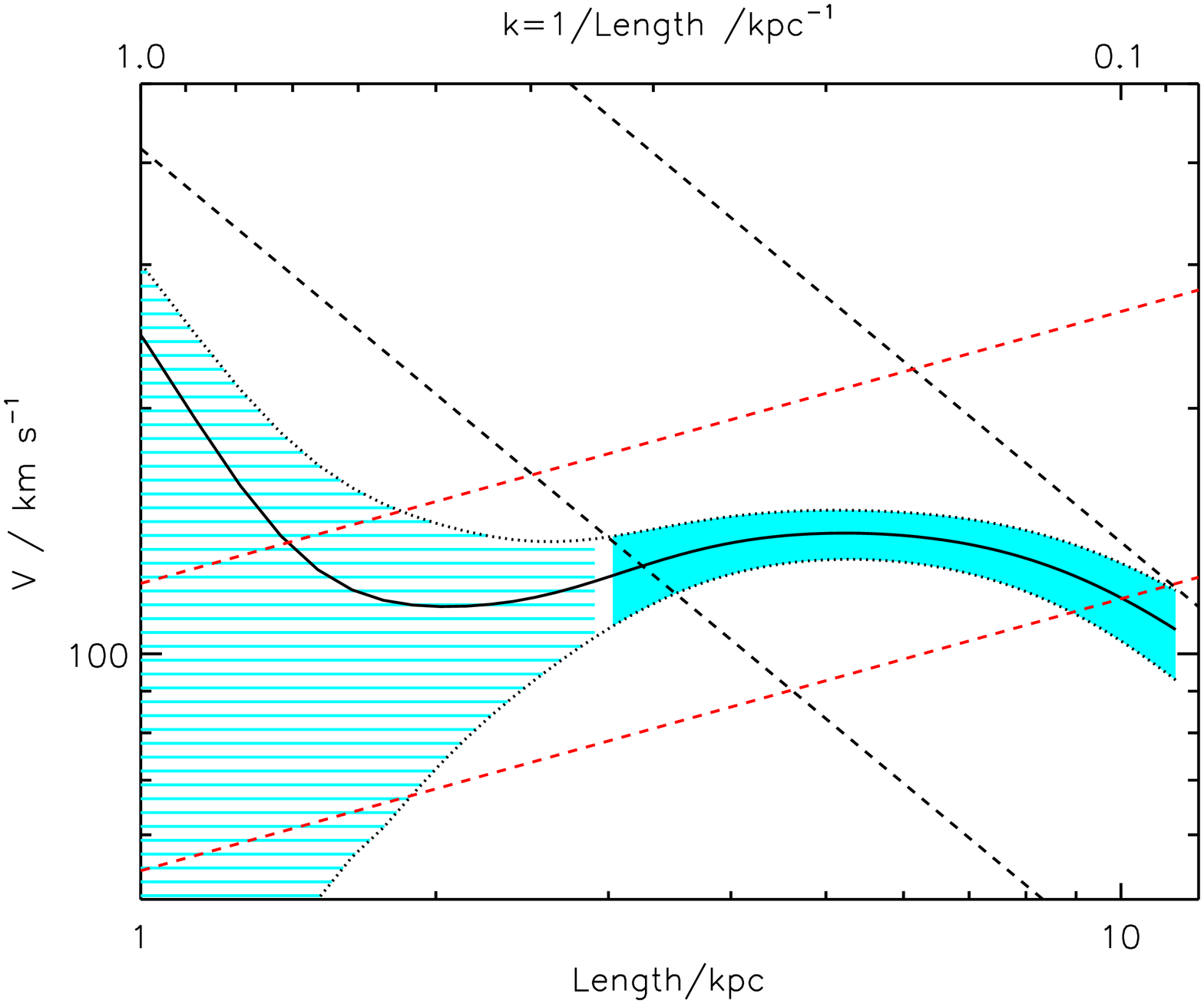,
        width=0.5\linewidth}
            }
    
      \caption{\emph{Left:} The solid black line shows the relation between $v$ 
and $l$ obtained by fixing the metal diffusion coefficient in equation 
\ref{diff_constraint} to the best fitting value found in \citet{Graham2006} for 
a radius at the midpoint of the sector. The dashed black lines show the allowed 
range for this relation when we take into account the the radial variation of 
the $D$ in the sector studied, and an uncertainty of a factor of 2 in the 
constant coefficient in equation \ref{diff_constraint}. The solid red line shows 
the relation obtained by equating the turbulent heating rate with the cooling 
rate in equation \ref{heating_constraint}. The dashed red lines show the range 
of this relation when the radial dependence of the gas density and cooling 
function are factored in, along with an uncertainty of a factor of 3 in the 
constant coefficient in equation \ref{heating_constraint}. The region where 
these two methods overlap provides a constraint on $v$ and $l$ for the gas 
motions \emph{Right:} Here we overplot the  constraints on $v$ and $l$ from the 
left panel onto those obtained obtained in Fig. \ref{velocitypowerspectra} by 
considering the power spectrum of excess surface brightness fluctuations.}
      \label{Rebusco_method_vl}
  \end{center}
\end{figure*}

Independent constraints on the velocities and spatial scales of the gas motions 
can be obtained by considering the metal diffusion coefficient, $D$, and the 
level of gas motions needed to offset cooling in the ICM. Following 
\citet{Rebusco2005} (who considered the Perseus cluster), under the assumption 
that metal diffusion is caused by stochastic gas motions on a length scale 
smaller than that distance from the centre of the cluster, $r$, we can write the 
diffusion coefficient in terms of the length scale, $l$ and velocity scale, $v$ 
of the gas motions 

\begin{eqnarray}
D \sim 0.11 v l
\label{diff_constraint}
\end{eqnarray} 

\citet{Graham2006} applied the stochastic model of metal transport developed by 
\citet{Rebusco2005} to the Centaurus cluster to obtain constraints on $D$. This 
method constrains $D$ by using the central abundance peak in galaxy clusters. 
The central abundance peak is likely produced by the stars in the BCG following 
the formation of the cluster, so the initial distribution of metals (when 
deposited) would be expected to follow the distribution of stars in the BCG. The 
observed metal peak is much broader than the light profile of the BCG, and if we 
assume the metals have been transported outwards by a stochastic process, 
constraints on $D$ can be made by comparing the two profiles. 

\citet{Graham2006} found that the best fitting diffusion coefficient had a 
radial dependence, decreasing with increasing radius. They found that $D$ could 
be described by 

\begin{eqnarray}
D \approx 4 \times 10^{28}  \left[  \frac{n_{H}(r)}{n_{H}(r_{0}) }\right] ^{2} 
{\rm cm^{2} s^{-1} }
\end{eqnarray}
where $n_{H}$ is the hydrogen density profile, and $r_{0}=25$kpc. Over the 
radial range of the sector we consider (23-43 kpc), $D$ decreases from 4.7 to 
1.5$\times 10^{28}$ cm$^{2}$ s$^{-1}$. We can then plot the constraints on this 
relation between $v$ and $l$, and this is shown with the black lines in the left 
hand panel of Fig. \ref{Rebusco_method_vl}.

Another relation between $v$ and $l$ can be obtained by requiring the heat 
dissipation rate of the gas motions to balance the cooling rate. From 
\citet{Rebusco2005}, the heating rate can be written as

\begin{eqnarray}
\Gamma_{diss} \sim 0.4 \rho v^{3} / l
\label{heating_rate_turbulence}
\end{eqnarray}
from which we obtain
\begin{eqnarray}
n(r)^2 \Lambda(T(r),A(r)) \sim 0.4 \rho v^{3} / l
\label{heating_constraint}
\end{eqnarray}
The red lines in the left hand panel of Fig. \ref{Rebusco_method_vl} show the 
constraints on $v$ and $l$ from this relation, where the range takes into 
account the range of gas density, temperature and metal abundance over the 
radial range of the sector considered. The constraints from equations 
\ref{diff_constraint} and \ref{heating_constraint} overlap in a region where 
they are both consistent, with heating balancing cooling and the metal diffusion 
coefficient agreeing with observations. 

In the right hand panel of Fig. \ref{Rebusco_method_vl}, we compare these 
constraints on $v$ and $l$ with those obtained earlier in Fig. 
\ref{velocitypowerspectra} by considering the power spectrum of excess surface 
brightness fluctuations. The two different methods appear to be in good 
agreement with one another.
 
\section{Conclusions}

We have examined the $\Delta$-variance power spectrum of the excess X-ray 
surface brightness fluctuations in a large sector to the west of the core of the 
Centaurus cluster, spanning a radial range between 23-43 kpc from the center of 
the cluster. When compared to the expected shape of the spectrum from Kolmogorov 
index fluctuations, the observed spectrum appears flatter. This is similar to 
what was found in \citet{SandersAWM7} for the fluctuation power spectrum of 
AWM7. The magnitude of the surface brightness fluctuations is roughly 8 percent 
on scales of 2 kpc. 

We have converted the 2D power spectrum into a 3D power spectrum and then found 
constraints on the characteristic velocity of the gas motions as a function of 
their spatial scaling following the methods of \citet{ChurazovComa} and 
\citet{Zhuravleva2014}. On scales between 4-10 kpc where the method is least 
sensitive to systematic and statistical errors, the one-component velocity of 
the gas motions is calculated to be in the range 100-150 km/s using this method. 
The turbulent heating rate from such gas motions is consistent with what is 
needed to balance the cooling rate, as was found for the Perseus and Virgo 
clusters in \citet{Zhuravleva2014}.

Independent constraints on the characteristic velocity and size scale of the gas 
motions have been obtained by measuring the diffusion coefficient needed to 
explain the present day distribution of metals, and invoking equilibrium between 
the turbulent dissipation rate and the gas cooling rate. This involves assuming 
that when the metals were first deposited by the stars in the BCG their 
abundance profile followed the light profile of the BCG. The constraints on $v$ 
and $l$ obtained using this method are in reasonable agreement with those 
obtained from the X-ray surface brightness fluctuations. 

It is possible that some fraction of the observed surface brightness fluctuations arise from sound waves rather than turbulence (Sanders et al. submitted). We investigate this in Appendix A, in which we compare the power spectrum from the whole sector with the power spectrum obtained by removing any features resembling sound waves. 

\section*{Acknowledgements}
SAW and ACF acknowledge support from ERC Advanced
Grant FEEDBACK. This
work is based on observations obtained with the \emph{Chandra} observatory, a 
NASA mission.
\bibliographystyle{mn2e}
\bibliography{Centaurus_fluctuations}

\appendix
\section[]{Linear features}
\label{sec:appendix}

\begin{figure*}
  \begin{center}
    \leavevmode
    \hbox{
      \epsfig{figure=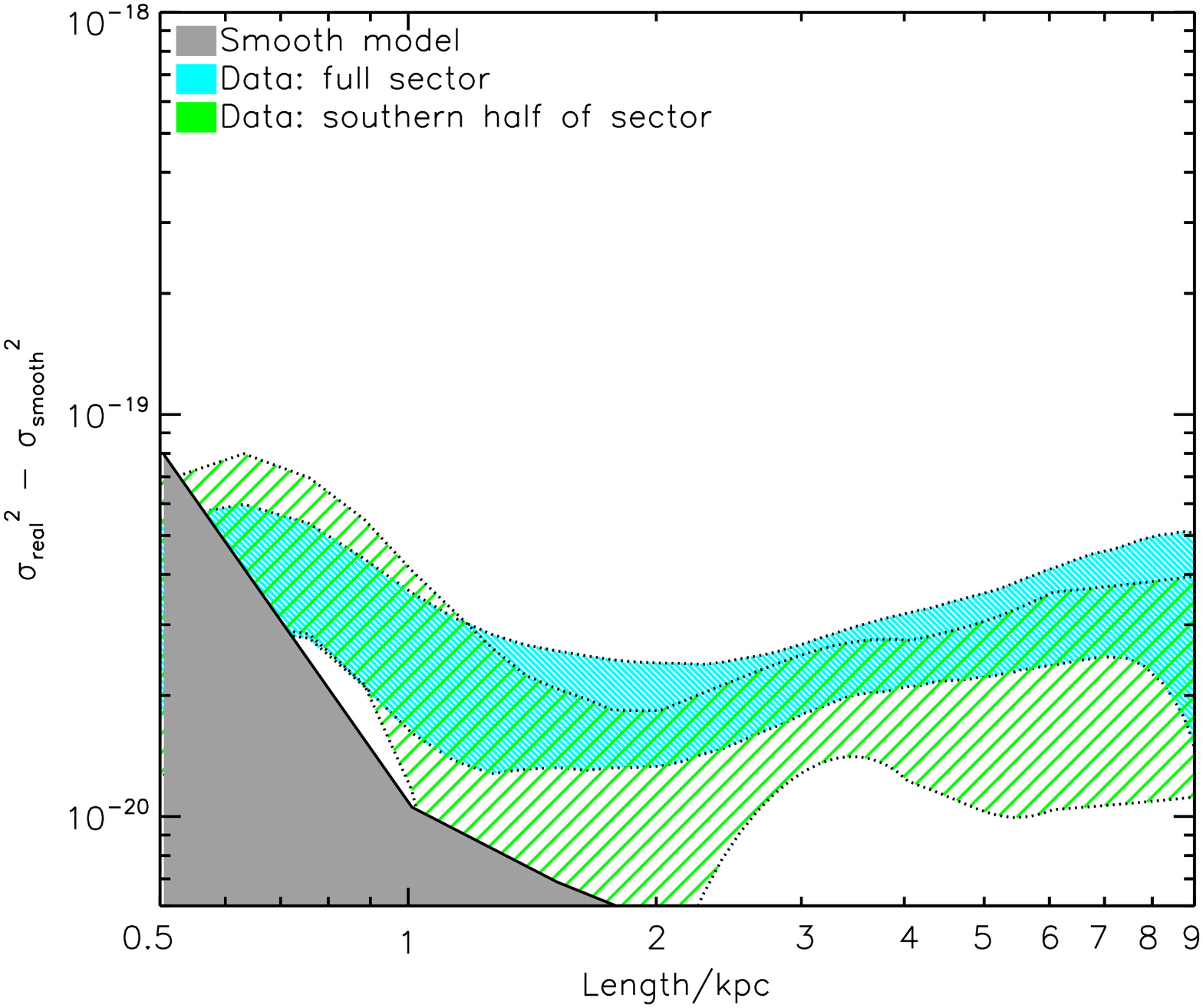,
        width=0.5\linewidth}
         \epsfig{figure=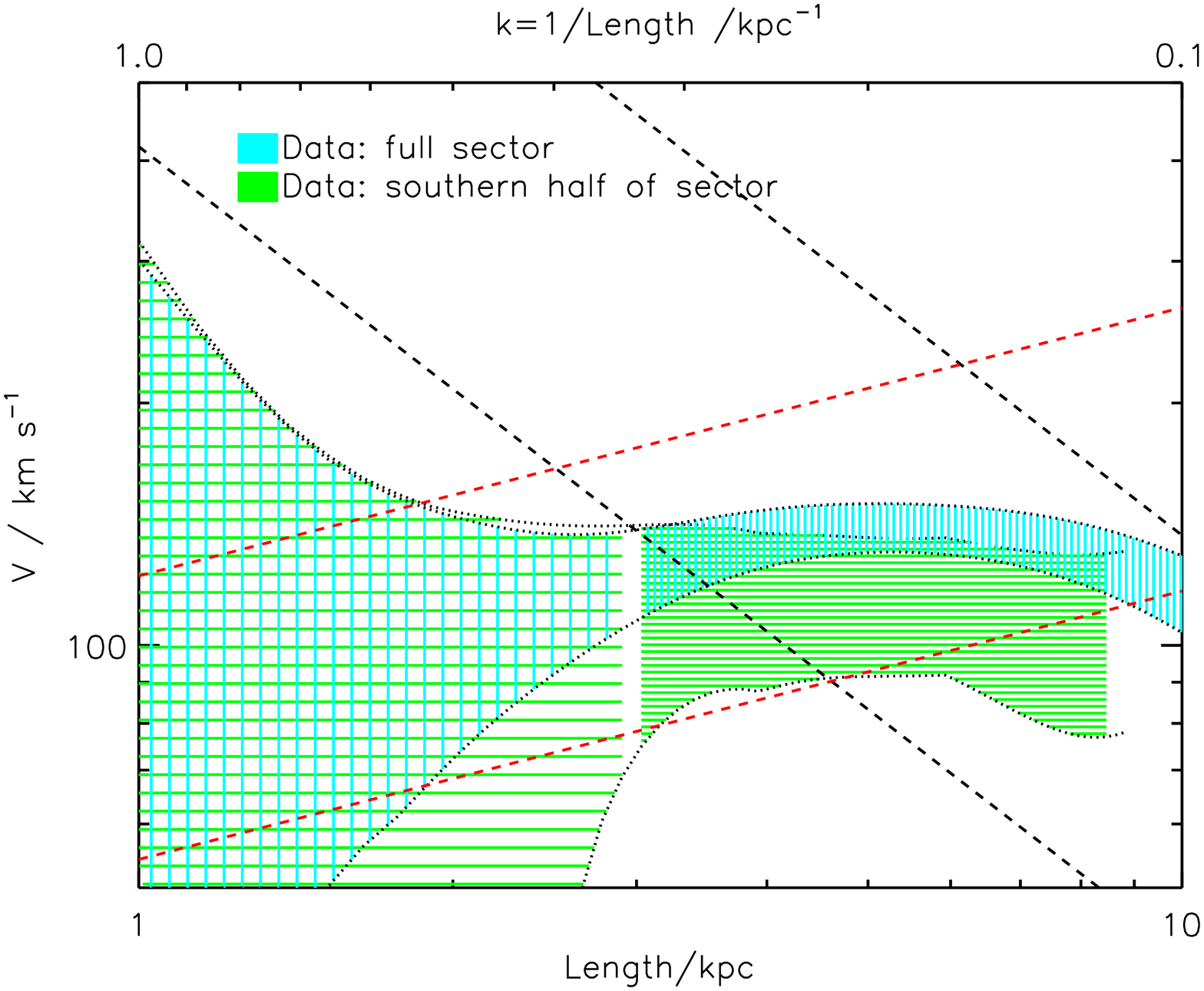,
        width=0.5\linewidth}
            }
    
      \caption{\emph{Left:}Comparing the power spectrum of the surface brightness fluctuations in the southern half of the sector away from any linear features (green line shaded region) with the power spectrum of the whole sector (cyan region).  \emph{Right:}Comparing the constraints on the one component velocity of the gas motions as a function of their length scale for the two regions. The black and red dashed lines show the constraints shown earlier in Fig. \ref{Rebusco_method_vl} obtained by considering metal diffusion and the heating-cooling balance. }
      \label{compare_to_south}
  \end{center}
\end{figure*}

Sanders et al. (submitted) have explored applying a Gaussian gradient magnitude (GGM) filter (similar to a Sobel filter) to the Chandra X-ray images, which emphasises gradients in the surface brightness. Sanders et al. (submitted) have identified several linear features in the northern part of the sector we have explored, which may be sound waves. This method is detailed in depth in in Sanders et al. (submitted). To investigate the affect of these, we repeated our analysis by excluding these features by examining just the southern half of the sector. The results are shown in Fig. \ref{compare_to_south}. In the left hand panel we compare the power spectrum of the whole sector (cyan region) with that of the southern half of the sector, which excludes the linear features (green region). We see that there is slightly less power in the southern half of the sector, though given the errors its power spectrum is consistent with that of the whole sector. This reduction in power lowers the velocities of the gas motions calculated from the surface brightness fluctuations, as shown in the right hand panel comparing the velocity constraints for the southern half of the sector with the whole sector. Given the errors the two are still consistent with one another, and also still consistent with the constraints on the velocity and length scales obtained by considering metal diffusion and the balance between heating and cooling (dashed black and red lines respectively).

\end{document}